# Cross-time functional connectivity analysis

Abbreviated title: cross-time functional connectivity

Ze Wang, PhD

ORCID: 0000-0002-8339-5567

Department of Diagnostic Radiology and Nuclear Medicine

University of Maryland School of Medicine

670 W. Baltimore St, Baltimore, MD 20201

ze.wang@som.umaryland.edu

Conflict of interest: none

Data availability statement: the HCP data is freely available from the HCP consortium.

Ethics approval statement: data reanalysis has been approved by IRB. Patient consent forms were obtained by HCP.

**Acknowledgements**

The research effort involved in this study was supported by NIH/NIA grants: R01AG060054, R01 AG070227, AG060054-02S1, and by the support of the University of Maryland, Baltimore, Institute for Clinical & Translational Research (ICTR) and the National Center for Advancing Translational Sciences (NCATS) Clinical Translational Science Award (CTSA) grant number 1UL1TR003098. Both imaging and behavior data were provided by the Human Connectome Project, WU-Minn Consortium (Principal Investigators: David Van Essen and Kamil Ugurbil; 1U54MH091657) funded by the 16 NIH Institutes and Centers that support the NIH Blueprint for Neuroscience Research; and by the McDonnell Center for Systems Neuroscience at Washington University in St. Louis. The authors thank the Human Connectome Project for open access to its data and thank Brigitte Pocta for editing the manuscript.

Abstract

A large body of literature has shown the substantial inter-regional functional connectivity in the mammal's brain. One important property remains un-studied is the cross-time interareal connection. This paper serves to provide a tool to characterize the cross-time functional connectivity. The method is extended from the temporal embedding based brain temporal coherence analysis. Both synthetic data and in-vivo data were used to evaluate the various properties of the cross-time functional connectivity matrix, which is also called the cross-regional temporal coherence matrix.

**Introduction**

Human brain is a highly self-organized (Willshaw 2006, Singer 2009, Haken 2012) and integrated network (Friston 2002). Underlying the functional self-organizations and integrations are the spatially and temporally correlated brain activities. Spatio-temporal correlations indicate brain interareal and cross-time brain interactions, which are fundamental to the network-wise brain behavior and the emergence of various brain functions. Numerous advances have been made in the past decades regarding the temporal brain coherence(Shaw 1984, Zang, He et al. 2007, Wang, Li et al. 2014) and interareal brain connectivity (Biswal, Yetkin et al. 1995, Gray and McCormick 1996, Calhoun, Adali et al. 2001, Varela, Lachaux et al. 2001, Buzsáki and Draguhn 2004, McIntosh, Chau et al. 2004, Fries 2005, Buzsaki 2006, Fries 2015, Wang 2019). Collectively, these studies have depicted an overall picture of the long-range temporal correlations and the long-range spatial correlations in the brain. Long-range temporal correlations mean that brain activity at one moment can influence future activities (Suckling, Wink et al. 2008, Wink, Bullmore et al. 2008, He, Zempel et al. 2010, He 2011) and the long-range spatial correlations suggest that local brain activity can affect remote brain regions (Friston 1995). The co-existence of both types of long-range correlations suggests a cross-time interareal interaction or equivalently a cross-time interareal functional connectivity, which have yet to be examined in the literature.

Two related measures published in the literature are the dynamic functional connectivity (DFC) (Hutchison, Womelsdorf et al. 2013) and the time lag of rsfMRI(Mitra, Snyder et al. 2014). The former is based on the zero-lagged time series in the same sliding window; the latter measures the relative time shift between two time series that makes the cross-correlation reach the maximum. In this paper, I proposed a temporal embedding-based method to characterize the cross-time interareal interactions or the cross-time function connectivity (CTFC) or the cross-regional temporal coherence (CTC) using resting state fMRI (rsfMRI). While these terms can be used interchangeably without introducing ambiguity, CTC was used in following text to reflect a connection to temporal coherence mapping which is under review in a separate paper.

CTC and CTC mapping (CTCM) can be considered as an extension of the temporal embedding-based brain entropy mapping work published in (Wang, Li et al. 2014). To consider the interactions between two time-lagged transit states from two regions, the transit state space, which is also called the phase space (Takens 1981), of each assessed region was reconstructed using temporal embedding. Temporal embedding is the process of using a short sliding window to extract the time segments: the so-called embedding vectors from the entire time series and then use them to form a high dimensional space with each embedding vector defining a sample of the phase space. This approach has been widely used to study the behavior of the time evolution trajectory of the system status and to predict future behavior. The window length is often set to be 2 or 3 in traditional temporal embedding, but was set to be a much bigger number like 30 in this paper in order to have sufficient samples for reliably estimating the correlation coefficient. After reconstructing the phase space for both regions, each embedding vector of the phase space of one brain region was compared to each embedding vector of the other brain region. The interareal embedding vector correlation was calculated and averaged across all possible pairs of comparisons. The positive and negative correlations were aggregated separately to examine the interareal correlation and anti-correlation separately. This separation allows us to examine the balance between correlation and anti-correlation (the C/A balance). Because correlation and anti-correlation may reflect excitation and inhibition respectively, the apparent C/A balance may provide a way to assess the macroscopic level excitation and inhibition.

Synthetic data and resting state functional MRI (rsfMRI) data were used to validate the CTC calculation method.

**Materials and Methods**

*Ethics statement*

Data acquisition and sharing have been approved by the HCP parent IRB. Written informed consent forms have been obtained from all subjects before any experiments. This study re-analyzed the HCP data and data Use Terms have been signed and approved by the WU-Minn HCP Consortium.

*Data included*

rsfMRI data, demographic data, and neurobehavior data from 862 healthy young subjects (age 22-37 yrs, male/female=398/464) were downloaded from HCP. The range of education years was 11-17 yrs with a mean and standard deviation of 14.86±1.82 yrs. The rsfMRI data used in this paper were the extended processed version released on July 21 2017. Each subject had four rsfMRI scans acquired with the same multi-band sequence(Moeller, Yacoub et al. 2010) but the readout directions differed: readout was from left to right (LR) for the $1^{st}$ and $3^{rd}$ scans and right to left (RL) for the other two scans. The purpose of acquiring different scans with opposite phase encoding directions was to compensate the long scan time induced image distortion. MR scanners all present field strength (B0) inhomogeneity, which causes signal distortion because of the imperfect excitation using the radiofrequency pulses that are tuned to the frequency determined by the ideal B0. While the B0 inhomogeneity caused distortions can be well corrected using two additionally acquired calibration scans using the opposite phase encoding directions: one is with LR and the other is with RL, HCP acquired two LR and two RL rsfMRI scans for the purpose of assessing the potential residual effects after the distortion correction and to assess the test-retest stability of rsfMRI measure. Each scan had 1200 timepoints. Other acquisition parameters for rsfMRI were: repetition time (TR)=720 ms, echo time (TE)=33.1ms, resolution

$2\times2\times2$ mm$^3$. The pre-processed rsfMRI data in the Montreal Neurological Institute (MNI) brain atlas space were downloaded from HCP and were smoothed with a Gaussian filter with full-width-at-half-maximum = 6mm to suppress the residual inter-subject brain structural difference after brain normalization and artifacts in rsfMRI data introduced by brain normalization.

*CTC calculation*

Fig. 1 illustrates the process of CTC calculation for two time series. Phase space reconstruction through temporal embedding is shown in the top and left panel along each of the two assessed time series. Denote one time series, for example, the rsfMRI time series of a brain voxel, by $x = [x_1, x_2, \ldots x_N]$, where N is the number of time points. The phase space of the underlying dynamic system can be reconstructed by a series of embedding vectors, each with m consecutive points extracted from x: $\mathbf{u}_i = [x_i, x_{i+1}, \ldots x_{i+m-1}]$, where $i = 1$ to N-m+1, m is the pre-defined embedding vector length. Note that the number of embedding vectors N-m+1 should be divided by the lag between temporally adjacent vectors if the lag is >1. For simplicity, the following derivations are based on a lag of 1 but can be easily extended to the case of lag>1. For the other time series y, the same embedding process can be used to obtain the corresponding embedding vectors. As illustrated by the lower panel of Fig. 1, CTC calculation is to first calculate the correlation coefficient matrix of all pairs of embedding vectors from x and y. For the simplicity of description, this matrix was named the CTC matrix in the following text.

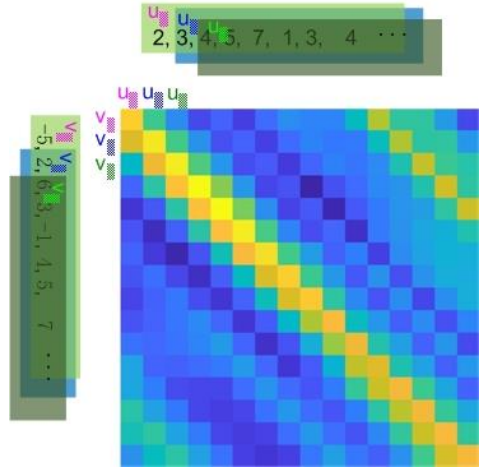

Figure 1. Illustration of CTC calculation. The left and upper panel show the moving window-based embedding vector extractions; the lower right panel shows the CTC matrix of those embedding vectors. Yellow means a correlation coefficient of 1; blue means negative correlations. $u_1$, $u_2$, $u_3$ and $v_1$, $v_2$, $v_3$ denote three embedding vectors from the two time series, respectively.

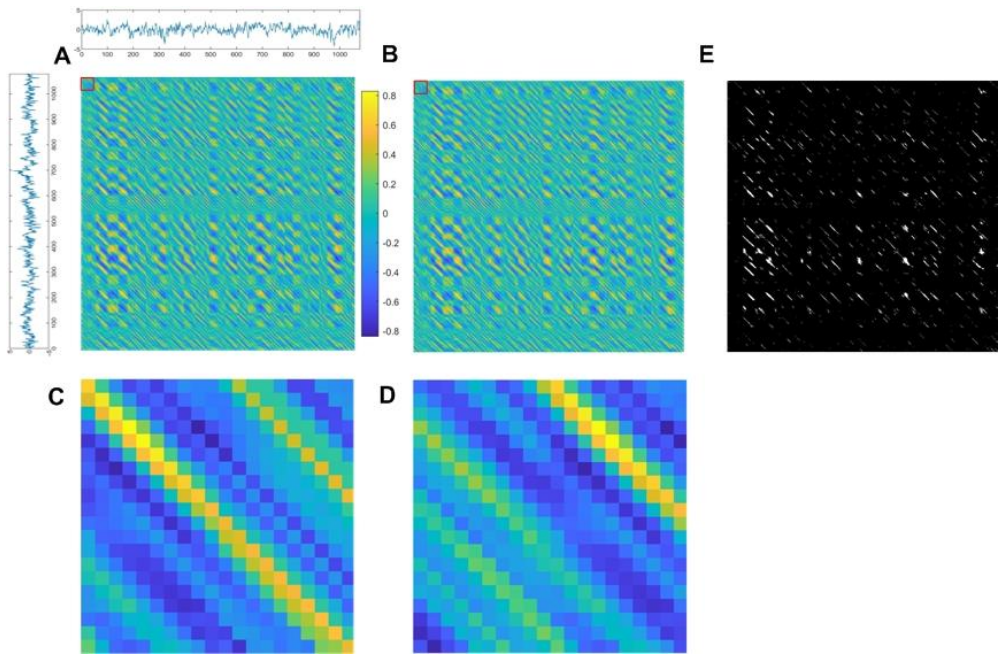

Fig. 2. CTC matrix of the embedding vectors of two rsfMRI time series extracted from a representative HCP rsfMRI scan in mPFC and the precuneus. A) The rsfMRI time series of mPFC

(on the left) and of precuneus (on the top), and the CTC matrix, B) the CTC matrix of mPFC and precuneus after the precuneus time series was intentionally delayed for 10 timepoints, C) and D) are the zoomed version of A and B at the location marked by the red squares. Compared to C, all coefficients in D were shifted to the right by 10 positions, which is as expected. E) the binarized CTC matrix with a threshold of 0.5.

Fig. 2 shows the CTC matrix between two brain regions: medial prefrontal cortex (mPFC) and precuneus. Fig. 2A and 2C are the CTC matrix and its zoomed version of the original rsfMRI time series from the two regions. Fig. 2B and 2D are the CTC matrix and the zoomed version after intentionally delaying the time series in the top panel of Fig. 2A by 10 time points. Clearly there was a directional CTC matrix shift along the orthogonal diagonal direction induced by the time shift. This shift also disturbed the balance between the mean CTC value of the upper triangles and that of the lower triangles of the CTC matrix. Accordingly, two empirical CTC metrics can be calculated as surrogate estimates of the lag between the two assessed time series. The first one is the off-diagonal position where the mean diagonal CTC value reaches the peak; the second is the ratio between the mean values of the upper triangle and the lower triangle of the CTC matrix. To validate these two metrics, four different synthetic time series were created, including: Gaussian noise (gau), a sinusoidal $\sin(0.01*pi*x)$ (sin), a random walk (randw), and a 1/f noise (1f). The two time series extracted from mPFC and precuneus/posterior cingulate cortex (PCC) in a representative HCP subject's preprocessed rsfMRI data (the first scan) was used to test the two lag measures too. For each time series, the center 1050 time points were extracted as the reference. Time lag was inserted by shifting the 1040 positions to the left (toward the starting point of the original time series) or to the right (toward the end of the time series). For each time-lagged time series, three methods were used to estimate the lag or the lag-related measures. The first one was the standard cross-correlation method where the lag was defined by the relative backward shift of the lagged time series that makes the back-shifted time series match the reference the most. Specifically, the lagged time series was shifted from -1/4 to 1/4 of the length (1040/4=260 here) in relative to the position of the reference. For each shift, the correlation between the reference and the shifted lagged time series was calculated. The shift where the correlation coefficient reaches the maximum was used as the global lag. This global method was previously used in (Mitra, Snyder et al. 2014). The second and the third ones were both based on the CTC matrix using the calculation process described above. The second method included the following steps: 1) CTC matrix calculation, 2) calculate the mean value of each diagonal of the CTC matrix, 3) find the diagonal whose mean value is the biggest out of the ones within the near neighbor of the main diagonal defined by -1/4 to 1/4 of the data length ($\pm$260). The offset of that diagonal in relative to the main diagonal is taken as the estimated lag. The third method was simply the ratio of mean positive correlation of the upper triangular matrix of CTC and that of the lower triangular matrix of CTC. Again, only the coefficient values within the area defined by the -1/4 and 1/4 data length off-diagonals were considered. 1/4 data length corresponds to π/2 and -π/2 to π/2 covers the cycle of the maximum frequency component of the assessed signal.

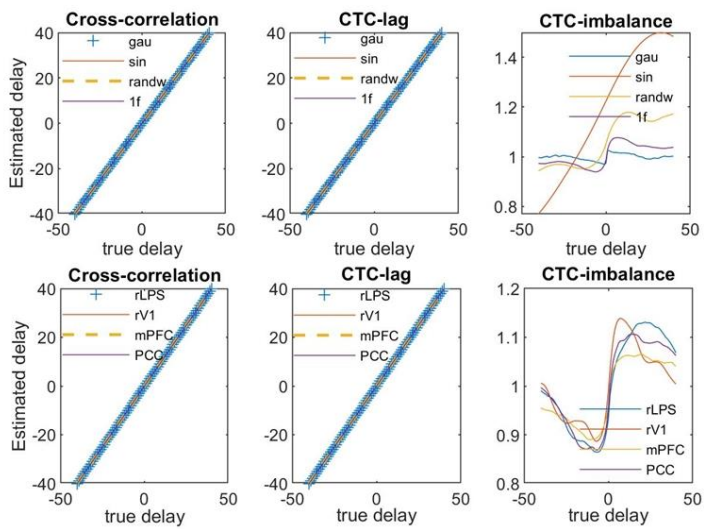

Fig. 3. Measures related to the lag between two time series calculated by different methods. Cross-correlation means the standard cross-correlation spectrum based method.

Fig. 3 shows that the conventional cross-correlation method and CTC method both accurately estimated the intentionally inserted time lag to the synthetic and in-vivo data. CTC imbalance monotonically increases with lag only when the lag was relatively small. A positive lag produced an imbalance ratio >1 and a negative lag resulted in a <1 imbalance ratio. Because the imbalance ratio only presents monotonical behavior in a very narrow delay range, it is limited for inferring large delays. Therefore, it was not included in the following experiments.

*Other CTC metrics*

In addition to time-lag, many other properties can be observed from the CTC matrix. Below, we listed a few basic metrics to characterize the most obvious properties of the CTC matrix.

(1) The first two are the cross-regional temporal coherence (CTC) and the cross-regional temporal anti-coherence (CTAC) which are calculated as the mean positive correlation coefficient and the mean negative correlation coefficient:

$$CTC = \frac{\sum_i^M cc_i}{M} \quad \text{for all } cc_i > 0$$

$$CTAC = \frac{\sum_i^M -cc_i}{M} \quad \text{for all } cc_i < 0$$

where M=(N-m+1)*(N-m+1) is the total number of elements of the CTC matrix.

(2) The third and the fourth are CTC_md and CTAC_md which are defined by CTC and CTAC at the main diagonal of the CTC matrix. Because the embedding vectors from the two compared time series are extracted at the same time window, CTC_md and CTAC_md are the same as the traditional sliding window-based FC except for the difference of separating positive FC from negative FC. Similar to CTC and CTAC, the introduction of

CTC_md and CTAC_md is to examine the cross-regional time-locked coherence and anti-coherence.

(3) The fifth and the sixth are the ratio between CTC and CTAC and the ratio between CTC_md and CTAC_md. These two measures indicate the balance between the coherence and anti-coherence (the C/A balance).

(4) The last three measures are based on the mean length of the continuous diagonal line segments of the binarized CTC matrix. The mean length of the continuous diagonal lines is defined in the recurrent plot analysis for measuring the divergency behavior of the dynamic system (Marwan, Romano et al. 2007). In this paper, we calculated this length for the positive and negative coefficients of the CTC matrix separately. As shown in Fig. 3, the CTC matrix can be binarized with a threshold to only retain the positions with stronger than the correlation strength cutoff. Isolated dots in the binarized matrix indicate rare transit states. In other words, they do not persist in time or they simply fluctuate too much. By contrast, the continuous diagonal line segments suggest that the system revisits the transit state represented by the embedding vectors at the corresponding coordinates of these segments many times. The length of those diagonal line segments provides an estimate of how long these transit states stick together. By hard-thresholding the CTC matrix with a positive threshold r, we can get a binarized positive correlation coefficient matrix (1 means CC>r, 0 means CC<=r). A negative threshold -r can be used to get a binarized negative CTC matrix (1 means CC<-r, 0 means CC>=-r). The following two measures can be used to measure the recurrence of coherent (positively correlated) and incoherent (negatively correlated) transit states, separately. Their ratio can be taken as another C/A balance indicator.

*MLP*: the mean length of the diagonal line segments of the binarized positive CTC matrix.

*MLN*: the mean length of the diagonal line segments of the binarized negative CTC matrix.

*CAR3=MLP/MLN:* the ratio of MLP to MLN

---

**Algorithm 1. CTC Mapping (CTCM)**

    input:
      x1 - time series with N timepoints
      x2 - time series with N timepoints
      w - embedding vector length
      r - correlation coefficient (CC) threshold
      g - gap between adjacent embedding vectors
    output:
      CTC - mean positive CC
      CTAC- absolute of the mean negative CC
      CAR1- CTC/CTAC
      CTC_lag – estimated delay between x1 and x2
      CTC_md - mean positive CC in the main diagonal
      CTAC_md - absolute of the mean negative CC in the main diagonal
      CAR2 – CTC_md/CTAC_md
      MLP - mean length of the diagonal lines of
          positive TCM matrix
      MLN - mean length of the diagonal lines of
          negative TCM matrix
      CAR3 - MLP/MLN

```
1   Nv=int((N-w+1)/g)  //#number of embedding vectors
2   DIA_E=Nv/4
3   max_dia=(Nv-DIA_E)*g
4   M=N-w-1
5   Totnum_cc = Nv*Nv
6   CTC=CTAC=0
7   MLP=MLN=0
8   Totnum_cc_band = Nv*Nv-DIA_E*(DIA_E+1)
9   for (l=0; l<Nv; l+=gap)
10    tlen_end=tlen-w-l
11    reset flags for detecting a diagonal line
12    seg_new=0; seg_s=0;   seg_e=0; mcc_d=0;
13    for(i=0; i < tlen_end; i=i+gap)
14       get the i-th embedding vector in x1 and the j-th embedding
15           vector in x2 (j= i+l*gap)
16       calculate their correlation cc
17       mcc_d +=cc;
18       if(cc>0) add cc to CTC
19       else    add cc to CTAC
20       if(cc>r)
21          if(seg_new==0)  // find a new diagonal line
22             // record the start point and set the flag
```

```
23              seg_s=i;   seg_e=i+1; seg_new=1;
24           else        // cotinuation of an existing line
25              seg_e+1 -> seg_e
26         Else
27            // set the end of the line if needed
28            if(seg_new==1)
29               seg_new=0; pseglen=seg_e-seg_s;
30               // exclude the isolated point
31               if(pseglen>1)
32                  MLP = MLP+pseglen;
33         if(cc<-thr)
34            if(nseg_new==0) // find a new diagonal line
35               nseg_s=i;   nseg_e=i+1; nseg_new=1;
36            Else
37               nseg_e+1 -> nseg_e;
38         Else
39            if(nseg_new==1)
40               nseg_new=0; nseglen=nseg_e-nseg_s;
41               if(nseglen>1)
42                  MLN = MLN+nseglen;
43      // force the last diagonal line to stop
44      if(seg_new==1)
45         seg_new=0;
46         pseglen=seg_e-seg_s;
47         if(pseglen>1)
48             MLP = MLP+pseglen;
49      if(nseg_new==1)
50         nseg_new=0;
51         nseglen=nseg_e-nseg_s;
52         if(nseglen>1)
53            MLN = MLN+ nseglen;
54      if l==0 set CTC_md =CTC/ Nv;
55              and CTAC_md =CTAC/ Nv;
56      if l<DIA_E  record CTC_lag: the location of the diagonal with the
57          maximum mean cc
58  Repeat lines 9-57 but change the l loop from 1 to Nv
             and calculate cc of the j-th embedding vector of x1 and the
             i-th embedding vector of x2
59   CTC =CTC/ Totnum_cc;
60   CTAC=CTAC/ Totnum_cc
61   CAR1=CTC/CTAC;
62   MLP =MLP/ Totnum_cc_band;
63   MLN =MLN/ Totnum_cc_band;
64   CAR2=CTC_md/CTAC_md;
```

65    CAR3=MLP/MLN;

---

*Algorithm and implementation*

Algorithm 1 describes a prototype algorithm of the CTC mapping (CTCM). Calculating the entire CTC matrix and then the parameters needs large computer memory and long computation time. Instead, I calculated the CTC properties on the fly. The main computation loop is for the delay between the embedding vectors, each from a different time series. In other words, the top loop of the algorithm is along the diagonal of the CTC matrix. "gap" is an integer here for increasing the time interval between adjacent embedding vectors. DIA_E is introduced to constrain the calculation of CTC_lag to be within -1/4 to 1/4 Nv around the main diagonal of the CTC matrix. This selection was based on the fact of that Nv corresponds to the cycle of the highest frequency component of the signal. Embedding vectors of a periodic signal with half cycle apart are the same. The algorithm was implemented in C++ using multiple threads-based parallel computing (Wang, Li et al. 2014, Wang 2021).

*Assessing the effects of w and r on the CTC parameters*

To evaluate the effects of w and r on temporal coherence characterization, we extracted mean time series from the first rsfMRI scan (the first LR scan) of 20 HCP subjects from the posterior cingulate cortex/precuneus (PCC/Pre) within a sphere with a radius of 6 mm in posterior cingulate/precuneus (PCC, coordinate in the MNI space: -5, -49, 40). This region-of-interest is widely used to define the default mode network (DMN) (Raichle, MacLeod et al. 2001) in standard seed-based FC analysis(Fox, Snyder et al. 2005). 20 1/f noise and gaussian noise were also generated. Mean time series were also extracted from two other regions: insular (ins) and motor cortex. CTCM was performed using the above algorithm with different w and r values: w varied from 30 to 90 with a step of 10; r varied from 0.2 to 0.6 with a step of 0.1. CTC, CTAC, CTC_md,

CTAC_md, CAR1, CAR2, MLP, MLN, and CAR3 were calculated. Analysis of variance (ANOVA) and paired-t test were used to statistically infer the effects of w when r was fixed or the effects of r when w was fixed.

*CTCM for the entire HCP dataset*

We then calculated the six parameters at each voxel for the 862 HCP subjects four rsfMRI data using a cluster of thousands of CPUs. The above mentioned PCC/Pre region-of-interest was used as the seed. The collections of each measure at all voxels form a corresponding CTC parametric map. For each subject, each of the six maps was averaged across the first LR and the first RL rsfMRI scans to minimize the potential effects of the phase encoding polarities. The resultant mean maps were then named by a prefix "M12". The same averaging process was performed for these parametric maps calculated from the second LR and the second RL rsfMRI scans. The resultant mean maps were then named by a prefix "M34". The M12 and M34 maps were used to test the test-retest stability of the different CTC measures.

*Statistical analysis.*

To find the potential biological or neurocognitive associations of resting brain TCM properties, we performed a series of voxelwise regression analyses. The biological measures included age and sex. Neurocognitive capability was measured by the Cognitive Function Composite score (TotCog) provided by HCP. The age unadjusted score was used with higher scores indicating higher levels of cognitive functioning. The score was calculated by averaging the normalized scores of each of the Fluid and Crystallized cognition measures, and then deriving scale scores based on this new distribution. Participant score normalization is in relative to those in the entire NIH Toolbox Normative Sample (18 and older), regardless of age or other variables. A score of 100 indicates performance that was at the national average and a score of 115 or 85, indicates performance one standard deviation above or below the national average.

For each type of CTC maps, a multiple regression model was estimated at each voxel using the map intensity at that voxel from all subjects as the independent variable. Dependent variables included age and sex for the age/sex association analyses. For TotCog, an independent regression model including age, sex, and TotCog as the covariates was estimated. All regression models were built and estimated using Nilearn (https://nilearn.github.io/). The voxelwise significance threshold for assessing each of the association analysis results was defined by $p<0.05$. Multiple comparison (across voxels) correction was performed with the family wise error (FWE) theory (Nichols and Hayasaka 2003) or the false discovery rate (FDR).

**Results**

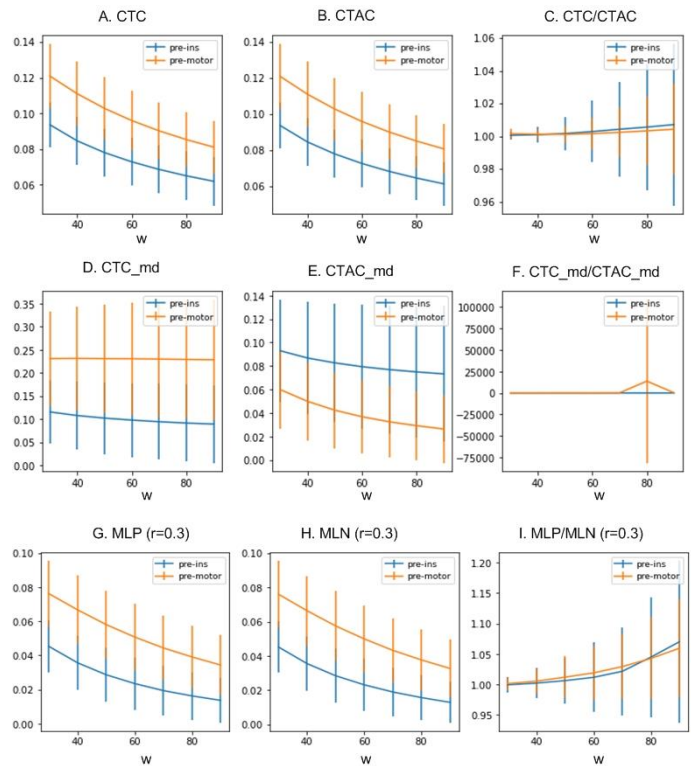

Fig. 3. CTC measures at different embedding vector length (w, the horizontal axis). A, B, C are independent of r. r=0.3 in G, H, and I. Error bars indicate STD of the measures from 50 different subjects. Pre, ins, and motor means precuneus, insula, motor cortex, respectively. pre-ins means CTC between precuneus and insular; pre-motor means CTC between precuneus and motor cortex.

*Effects of w on CTCM results*

Fig. 3 shows the evaluation results of CTCM with different parameters, w, and r, for CTC between the seed: precuneus (pre) and each of the two target regions: insular (ins) and motor cortex (motor). ANOVA was used to assess the effects of different w values on CTC parameters. For each of the two regions pairs: pre-ins and pre-motor, both CTC (Fig. 3A) and CTAC (Fig. 3B) decreased (p<4e-33, p<1.6e-38) with w (p<1.6e-38). CAR1 (Fig. 3C) did not show significant change when w changed (p>0.88). CTC_md (Fig. 3D), CTAC_md (Fig. 3E), and CAR2 (Fig. 3F) did not show significant change when w changed except for pre-motor CTAC_md, which significantly increased with w (p=2.5e-7). Figs 3G-3I show MLP, MLN, and CAR3 for r=0.3. Similar results were found for other r values. Both MLP (Fig. 3G) and MLN (Fig. 3H) decreases with w (p<2e-27) but CAR3 (MLP/MLN, Fig. 3I) significantly increases with w for both pairs of time series (p<3.8e-6). For all assessed w and r, CTC, CTAC, CTC_md, CTAC_md, MLP, MLN are significantly different between insula and motor cortex when CTC was calculated against precuneus (pre-ins and pre-motor) (p<0.00014). CAR1, CAR2, and CAR3 of precuneus CTC were significantly different between insula and motor only at w=30 (p=0.018, p=0.041, p=0.038, respectively. r=0.6 for CAR3).

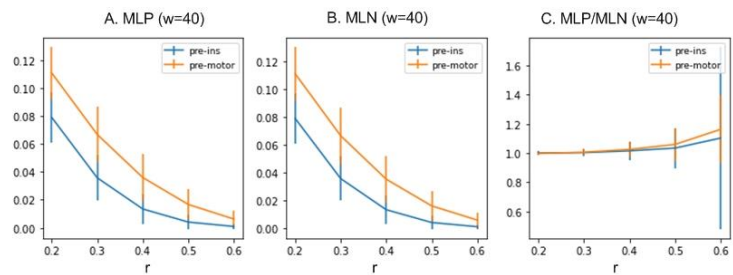

Fig. 4. The mean length of the continous diagonal segments of the precuneus CTC matrix binarized by different cutoffs (r). Error bars indicate STD of the measures calculated from 50 different subjects. Pre, ins, and motor means precuneus, insula, motor cortex, respectively. pre-ins means CTC between precuneus and insular; pre-motor means CTC between precuneus and motor cortex.

*Effects of r on MLP, MLN, and CAR3*

Fig. 4 shows the effects of r on MLP, MLN, and CAR3. Both MLP and MLN decrease with r ($p<1.4e-59$, one-way ANOVA). CAR3 did not show significant r effects for pre-ins CTC but had significant r effects in pre-motor CTC ($p<1.42e-8$, one-way ANOVA). MLP significantly differed between pre-ins and pre-motor CTC (paired t-test, $p<2.9e-5$) for all evaluated r values. MLN significantly differed between pre-ins and pre-motor CTC (paired t-test, $p<0.00014$). CAR3 of pre-ins and pre-motor did not differ significantly for all r and w except for r=0.6 and w=30 ($p=0.038$).

*Mean CTC parametric maps*

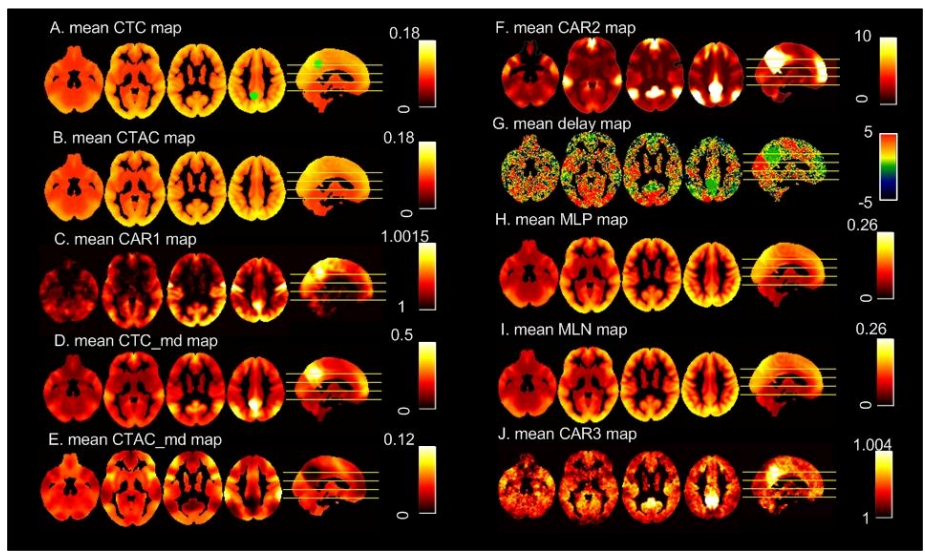

Fig. 5. Mean CTC maps from 862 young healthy adults. The green spot in A marks the position of the PCC/Pre seed.

Fig 5 shows the mean PCC/Pre CTC maps of the 862 HCP young healthy subjects' first and second rsfMRI scan (the mean of the 862 subjects' M12 CTC maps). The mean of the M34 maps were almost identical to the mean of the M12 maps, so they were not shown here. CTC, CTAC, MLP, and MLN (Figs 5A, 5B, 5H, 5I) showed similar image contrast between grey matter and white matter (note that part of white matter has been masked out during CTCM to save some computation time). Subcortical regions and cerebellum showed lower value than cortex. None of the four maps showed clear patterns in the DMN. CAR1 (Fig. 5C) shows higher intensity in precuneus/PCC, visual cortex, motor cortex, and parietal cortex. Precuneus/PCC and parietal cortex are within the standard DMN network. A typical DMN pattern was found in the CTC_md map (Fig. 5D) consisting the regions with high intensity in medial prefrontal cortex, temporal cortex, parietal cortex, and precuneus. Fig. 5E is the mean of CTAC_md. Compared to CTC_md, CTAC_md shows high intensity in the frontal-parietal regions which are known as the dorsal attention network (DAN) or the executive control network (Shulman, Corbetta et al. 1997, Damoiseaux, Rombouts et al. 2006). Meanwhile, CTAC_md shows the lowest intensity in the DMN. CAR2 (Fig. 5F) shows an enhanced contrast between the DMN and the DAN with the former showing higher than average intensity and the latter showing lower than average intensity. CTC_lag (Fig. 5G) reveals no delay to PCC/Pre (the seed) in the DMN but a delay about 5 TRs (3.6 secs) to PCC/Pre in the visual cortex and motor network. CAR3 (Fig. 5J) showed enhanced contrast in precuneus and temporal cortex.

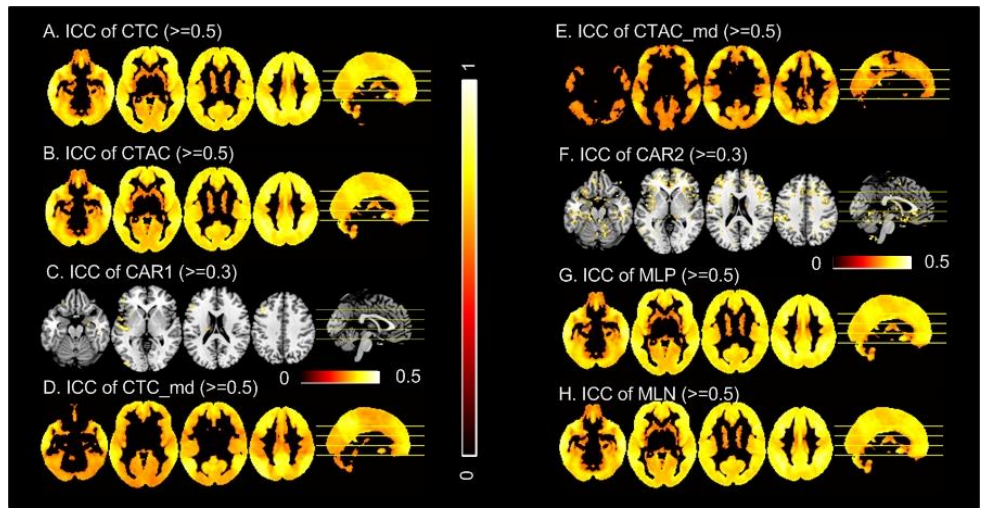

Fig. 6. ICC of CTC parameter maps.

*Test-retest stability of CTC parameters*

The M12 (mean of the results derived from the first two rsfMRI scans) and M34 (mean of the results derived from the last two rsfMRI scans) CTC maps were used to assess the test-retest reliability of different CTC parameters. Fig. 6 shows the results thresholded at 0.3 (moderate reliability) and 0.5 (high reliability). CTC, CTAC, MLP, and MLN show high reliability in nearly the whole brain (note that CSF and part of the white matter was excluded from computation). CTC_md and CTAC_md were reliable in cortex and most part of cerebellum. CAR1 and CAR2 showed low to moderate reliability in a very small part of brain. CTC_lag (CTC delay), CAR3 was not reliable even with a low cutoff of ICC=0.3 and were not shown in Fig. 6.

*Age effects on PCC/Pre CTC parameters*

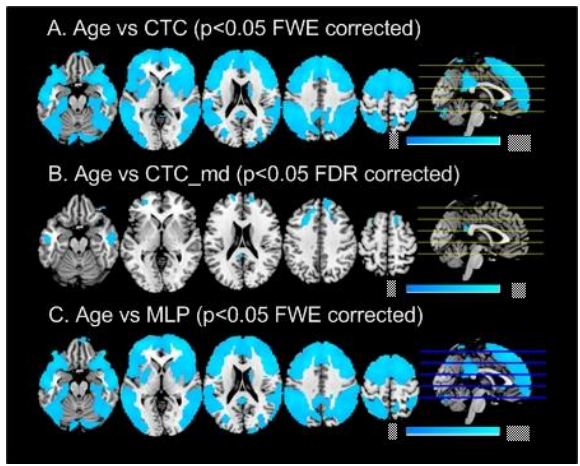

Fig. 7. Age effects on the CTC parameters.

Fig. 7 shows the correlations between PCC/Pre CTC parameters and age. CTC and MLP exhibit significant (FWE corrected) negative age effects in the frontal cortex, parietal cortex, temporal cortex, and part of visual cortex (fusiform and V2). CTC_md (Fig. 7B) presents statistically significant (p<0.05 FDR corrected. Cluster size>150) negative age effects only in the medial prefrontal cortex, middle temporal cortex, and a small section of PCC. CTAC had nearly the same age effects as those in CTC (Fig. 7A); CTAC_md had nearly the same age effects as CTC_md; MLN had nearly the same age effects as MLP. Therefore, their age effects were not displayed in Fig. 7. CAR1 did not show any significant age effects.

*Neurocognitive correlates of PCC/Pre CTC parameters*

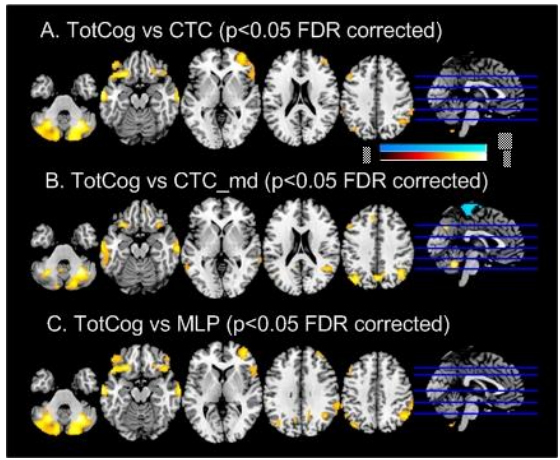

Fig. 8. Associations of CTC parameters to total cognitive scores.

Fig. 8 shows the correlations between the total cognitive scores and PCC/Pre CTC metrics. The results of CTAC, CTAC_md, and MLN are nearly the same as those of CTC (Fig. 8A), CTC_md (Fig. 8B), and MLP (Fig. 8C), respectively, and were not displayed in this figure. CTC (Fig. 8A) showed similar cognitive correlation patterns to those of MLP. The positive correlations were located in ventro-medial prefrontal cortex, dorso-lateral prefrontal cortex, middle temporal lobe, cerebellum, precuneus, and angular gyrus (lateral parietal cortex). CTC_md showed spatially less distributed correlation patterns in those regions. Interestingly, CTC_md was negatively correlated with total cognitive scores in the primary motor cortex. Other CTC measures including CTC delay, CAR1, CAR2, and CAR3 did not show significantly correlations with the total cognitive scores under the statistical significance cutoff of $p<0.05$ (FDR corrected) and a cluster size >150.

**Discussion**

The cross time functional connectivity or the cross-regional temporal coherence was investigated using rsfMRI data in this paper. Similar to the seed-based FC, CTC is in relative to a seed. Different from the traditional FC, CTC is defined by the correlations between time-lagged sub time series of the seed and the target region. By varying the time lag, a CTC matrix was calculated, and several collective metrics were then proposed to characterize the properties of the CTC matrix. Using intentionally delayed data, the maximum diagonal mean of the CTC matrix was shown to be sensitive to time lag between the seed and the target time series. The location of the maximum diagonal correctly reflected the time lag and was the same as the lag measured by the cross-correlation method (Mitra, Snyder et al. 2014). The ratio of the mean of the upper triangular CTC and the mean of lower triangular CTC monotonically increased with the lag but in a very limited range.  By considering the positive and negative correlations separately, CTC (the mean positive correlation coefficients in the CTC matrix) and CTAC (the mean negative correlation coefficients in the CTC matrix) were proposed to measure the cross-regional temporal coherence and cross-regional temporal anti-coherence separately. CAR1 was used to quantify the balance between CTC and CTAC. By constraining CTC and CTAC to the main diagonal of the CTC matrix, CTC_md and CTAC_md were defined to characterize the simultaneous cross-regional temporal coherence and anti-coherence, similar to the DFC (dynamic functional connectivity) analysis but differ from that by considering the correlation and anti-correlation separately. CAR2 was used to measure the balance between CTC_md and CTAC_md. Because CTC depends on the window length w for extracting the temporary segment of the time series, different w's were used to evaluate the effect of w on the CTC metrics and their difference between different time series when the same seed region was used. The experiment results showed that both CTC and CTAC decreases with w. The slop of CTC and CTAC becomes smaller when w increases. These results are caused by the addition of more data variations when w increases, which reduces the overall correlation coefficient magnitude. When w further increases, the extra variations due to the insertion of more data become less effective for the correlation coefficient. CTC_md and CTAC_md did not show significant changes with w except for the PCC/Pre vs Motor cortex CTAC_md. This w effect difference between CTC/CTAC and CTC_md/CTAC_md may be caused by the underlying number of coefficients used to calculate the measures, which is about ½ $N_v$ times bigger in the CTC and CTAC than in the CTC_md/CTAC_md, where $N_v$ is the total number of embedding vectors in one time series. CAR1, CAR2, CAR3 were insensitive to w. MLP and MLN were used to quantify the average time for the target process to stay coherent with the seed process. These two measures are physiologically meaningful as two interacting processes likely to have transit interactions for a consecutive time rather than in some singular or isolated time. Both MLP and MLN depends on the correlation coefficient cutoff. The evaluations results showed that MLP and MLN both decrease with w and r. This is because the correlation coefficients tend to decrease with w. Using the same r threshold, the number of suprathreshold coefficients decrease with w, so does the mean adherence time length (MLP and MLN). Similarly, when r increases, the number of suprathreshold coefficients decreases, so do MLP and MLN. The minor increase of CAR3 (the ratior of MLP and MLN) with respect to w means that the effects of w on MLN are more than those on MLP. For all w and r, CTC, CTAC, CTC_md, CTAC_md, MLP, and MLN all differed between the time series of insula and motor cortex in relative to PCC/Pre. CAR1, CAR2, and CAR3 significantly differed between insula and motor cortex only at w=30.

The whole brain CTC parameter maps were then calculated using the HCP rsfMRI data. The seed was set to be the PCC/Pre, which is widely used for assessing the DMN. CTC, CTAC, MLP, and MLN show similar image contrast with relatively high intensity in the frontal parietal network, visual cortex, and temporal lobe. This pattern differed from the one shown in CTC_md which was similar to the typical DMN and can be explained by the fact of that CTC_md is equivalent to the mean DFC. CTAC_md presented lower intensity in DMN because the PCC/Pre is well known to be positively correlated with the other parts of the DMN; CTAC_md showed relatively higher intensity in the DAN which is known to be negatively correlated with PCC/Pre. The CAR1 maps show that for the PCC/Pre CTC, the positive CTC differed from the negative CTC the most in the posterior part of the DMN (precuneus and parietal cortex) and the sensorimotor network (visual cortex and motor cortex) though the difference was considered minor (about 0.1%). CAR2 showed an enhanced DMN pattern (higher intensity) and DAN pattern (very lower intensity). CTC delay to

PCC/Pre was zero in the DMN and greater than 0 in sensorimotor network (visual cortex and motor cortex). The CAR3 map showed that MLP is greater than MLN by >0.2% in PCC/Pre and temporal cortex. Together, these results showed that regions within the same network as the seed are more positively connected across time; positive and negative interareal interactions are well balanced in most of the brain. The biggest positive to negative interaction imbalance was shown in the positive network and negative network as defined by the seed; within-network voxels had no delay to the seed; cortical regions show stronger cross-time temporal coherence and anti-coherence with the cortical seed than subcortical regions and white matter; cortical regions coherently or anti-coherently stay with the cortical seed for longer than than subcortical regions and white matter; based on the cross-time CTC measures, including CTC, CTAC, MLP, and MLN, we can see that nearly the entire cortical regions show cross-time coherence and anti-coherence to the seed.

CTC, CTAC, MLP, and MLN showed high test-retest stability in nearly the entire grey matter and white matter. The highest stability of these four maps was in the DMN and DAN. CTC_md and CTAC_md showed high test-retest stability mainly in grey matter. The highest stability was in DMN in CTC_md and was in DAN in CTAC_md. CAR1 and CAR2 only showed low reliablity in very small regions. CTC delay and CAR3 was not reliable in the entire brain.

The mean cross-time CTC measures: CTC, CTAC, MLP, and MLN presented signficant age effects (negative correlations) in most of the cortex and cerebellum except for motor network, while the time-locked mean CTC measures: CTC_md and CTAC_md showed age effects (negative correlations) only in the DMN (medial prefrontal cortex, temporal cortex, cerebellum, and precuneus). These results were consistent with the age effects of seed-based FC(Andrews-Hanna, Snyder et al. 2007, Biswal, Mennes et al. 2010, Jones, Machulda et al. 2011, Mevel, Landeau et al. 2013). Age effects on other measures (CTC delay, CAR1, CAR2, CAR3) were not examined since their test-retest reliability is low. Both the mean cross-time CTC measures (CTC, CTAC, MLP, and MLN) and time-locked CTC measures (CTC_md and CTAC_md) showed age-independent positive correlations to the total cognitive scores, which is consistent with the cognitive impairment-related DMN FC reduction reported in (Lucas-Jiménez, Ojeda et al. 2016).

Several limitations must be noted. First, as a proof-of-concept paper, only one seed was assessed in this paper. Second, CTC vs the total cognitive scores rather than a specific domain function was examined. This is because the seed was chosen to be the PCC/Pre which is often used to assess the DMN and DMN is often believed to be non-specific to domain functions(Raichle, MacLeod et al. 2001, Raichle and Gusnard 2005, Raichle and Snyder 2007, Raichle 2015).

These evaluations suggest the six CTC measures:CTC, CTAC, CTC_md, CTAC_md, MLP, and MLN as biologically reliable and cognitively meaningful resting state brain measures. Other measures are methodologically sound but are not reliable, suggesting large dynamic variabilities. Overall, the resting healthy brain presents profound cross-time interareal coherence and anti-coherence; transit brain state tends to stay coherent or anticoherent for a longer time in the cortical area than in the subcortical regions and white matter; coherence and anticoherence are well balanced in the brain with minor to moderate fluctuations.